\documentclass[pra,showpacs]{revtex4}
\usepackage{amsmath}
\usepackage{graphicx}
\usepackage{makeidx}
\usepackage{subfigure}

\begin{document}

\title{Localized matter-waves patterns with attractive interaction in
rotating potentials}
\author{Hidetsugu Sakaguchi$^{1}$ and Boris A. Malomed$^{2}$}
\affiliation{$^{1}$Department of Applied Science for Electronics and Materials,\\
Interdisciplinary Graduate School of Engineering Sciences,\\
Kyushu University, Kasuga, Fukuoka 816-8580, Japan\\
$^{2}$Department of Interdisciplinary Studies,\\
Faculty of Engineering, Tel Aviv University,\\
Tel Aviv 69978, Israel}

\begin{abstract}
We consider a two-dimensional (2D) model of a rotating attractive
Bose-Einstein condensate (BEC), trapped in an external potential.
First, an harmonic potential with the critical strength is
considered, which generates quasi-solitons at the lowest Landau
level (LLL). We describe a family of the LLL quasi-solitons using
both numerical method and a variational approximation (VA), which
are in good agreement with each other. We demonstrate that kicking
the LLL mode or applying a ramp potential sets it in the Larmor
(cyclotron) motion, that can also be accurately modeled by the VA.
Collisions between two such moving modes may be elastic or
inelastic, depending on their total norm. If an additional
confining potential is applied along with the ramp, it creates a
stationary edge state. Applying a kick to the edge state in the
direction of the ramp gives rise to a skipping motion in the
perpendicular direction. These regimes may be interpreted as the
Hall effect for the quasi-solitons. Next, we consider the
condensate trapped in an axisymmetric quartic potential. Three
species of localized states and their stability regions are
identified, \textit{viz}., vortices with arbitrary topological
charge $m$, ``crescents" (mixed-vorticity states), and strongly
localized center-of-mass (c.m.) states, alias quasi-solitons,
shifted off the rotation pivot. These results are similar to those
reported before for the model with a combined quadratic-quartic
trap. Stable pairs of c.m. states set at diametrically opposite
points are found too. We present a VA which provides for an
accurate description of vortices with all values of $m$, and of
the c.m. states. We also demonstrate that kicking them in the
azimuthal direction sets the quasi-solitons in
\textit{epitrochoidal} motion (which is also accurately predicted
by the VA), collisions between them being elastic.
\end{abstract}

\maketitle

\section{Introduction}

Formation of vortices is a well-known manifestation of
superfluidity in Bose-Einstein condensates (BECs) -- in
particular, in those which form effectively two-dimensional
(``pancake") configurations in appropriately designed trapping
potentials \cite{book}. If a condensate with repulsive
interactions between atoms, which is confined in a nearly 2D layer
by a 2D harmonic (quadratic) potential with trapping frequency
$\omega $, is set in rotation with frequency $\Omega $, formation
of a multi-vortex lattice is observed in the experiment
\cite{Ketterle}. The stability of such lattices is limited to
$\Omega <\omega $, as otherwise the centrifugal force empties the
region in the center of the trap \cite{omega=Omega}. Close to the
instability threshold, formation of a metastable state in the form
of a giant vortex, with topological charge $m\sim 50$, was
observed \cite{giant}. On the other hand, it was proposed
theoretically \cite{quartic-prediction} and implemented in the
experiment \cite{quartic-exp} that the instability at $\Omega \geq
\omega $ can be eliminated if the trapping potential is steeper
than harmonic, the simplest possibility being to add a quartic
term to it [in the critical case of $\Omega =\omega $, the
linearized version of the respective 2D Gross-Pitaevskii equation
(GPE) is tantamount to the Schr\"{o}dinger equation for a charged
particle in the uniform magnetic field, which gives rise to the
Landau levels, see Eq. (\ref{LLL}) below]. In many theoretical
works, it has been demonstrated that self-repulsive condensates
can form stable vortices with a multiple topological charge,
$m>1$, in 2D anharmonic traps \cite{multiple}.

Dynamics of vortices in BEC with attraction between atoms is different -- in
particular, due to the possibility of the collapse in self-attractive media
\cite{collapse}. The stability of 2D vortices confined by the harmonic
potential was studied in detail \cite{Dumitru}. It was demonstrated that,
prior to the onset of the collapse, the vortex with $m=1$ is destabilized by
azimuthal perturbations that split it into mobile localized objects
resembling fundamental solitons, if the norm of the vortex exceeds a certain
critical value. Vortices with $m\geq 2$ are completely unstable in the same
setting.

It was also predicted that, under the action of the rotation, BEC
with the intrinsic attraction can break the 2D axial symmetry by
self-trapping into quasi-soliton objects, alias ``center-of mass"
(c.m.)
states, characterized by an offset of the c.m. from the rotation pivot \cite%
{cm}. It was concluded that the anharmonicity of the trapping potential is
necessary for the stability of the c.m. states \cite{cm-anharmonic}.
Therefore, the theoretical study of rotating attractive condensates trapped
in quadratic-quartic radial potentials has drawn attention. Phase diagrams
of this model were investigated in detail, both in the mean-field
approximation [i.e., using the GPE and its linearization for small
perturbations, in the form of the Bogoliubov - de Gennes equations] \cite%
{mean-field}, and by means of a numerical diagonalization of the many-body
bosonic Hamiltonian \cite{many-body}, both approaches producing similar
results. Three types of stable localized patterns were identified in these
studies: vortices with topological charge $m=0,1,2,3,...$; crescent-shaped
states with a broken axial symmetry, that may be realized, for instance, as
a superposition of vortices with $m=2,3,4$; and the c.m. states shifted from
the rotation pivot. The transition between the crescents and c.m. states
which feature stronger localization is gradual. It is also relevant to
mention that crescents built similar to those reported in Refs. \cite%
{cm-anharmonic} and \cite{mean-field} can be made stable in a completely
different model, viz., a quasi-linear 2D equation with the harmonic trap
whose strength is proportional to the total norm of the configuration \cite%
{He} (the so-called ``accessible-soliton" model, which is relevant
to nonlinear optics \cite{Alan}).

The previous analysis of these states was carried out, in the framework of
the GPE, in a numerical form only. One of purposes of the present work is to
demonstrate that the entire family of vortices, with all values of $m$, and
the well-localized c.m. states can be predicted, in an accurate form, by a
simple variational approximation (VA) (this method was first applied to BEC
in Ref. \cite{VPG}; for a general review of variational techniques for
solitons, see Ref. \cite{Progress}). In addition, we find stable pairs of
c.m. states set at diametrically opposite points. Another objective of this
work is to consider the motion of c.m. states and collisions between them
(which turn out to be elastic). The motion is also accurately described by
the VA. To focus on effects of anharmonic traps, in that part of the work we
consider the GPE with the quartic radial potential only, which may be
implemented in the experiment \cite{quartic-exp}.

Another setting considered in this work is the above-mentioned critical case
of the quadratic axisymmetric trap with $\omega =\Omega $. In this case, we
demonstrate that quasi-solitons can be found in the lowest Landau level
(LLL). Both quiescent and moving LLL modes are very accurately described by
an appropriately modified VA. The motion, of the Larmor (cyclotron) type, is
imposed on them by the application of a kick, or by the action of a 1D ramp
potential. In addition, we consider the situation when the ramp acts in a
combination with a 1D quartic potential. In the latter case, we find edge
states of the nonlinear LLL modes, and also study their motion induced by
the kick, which may be interpreted as an effective Hall effect for the
quasi-solitons.

The paper is structured as follows. The underlying two-dimensional GPE and a
brief description of the numerical method employed to look for stationary
solutions (which is based on the integration in imaginary time) are
presented in Section II. In Section III, we consider the quasi-solitons of
the LLL type in the critical model with the quadratic axisymmetric confining
potential, while the model with the quartic potential is dealt with in
Section IV. The paper is concluded by Section V.

\section{The model and numerical methods}

The mean-field approximation for the 2D condensate trapped in potential $%
U(x,y)$ and rotating at angular velocity $\Omega $ is based on the GPE for
the single-atom wave function, $\psi (x,y,t)$. In the scaled form, the
equation, written in rotating coordinates $x$ and $y$, takes the well-known
form \cite{cm}-\cite{mean-field}:
\begin{equation}
i\frac{\partial \psi }{\partial t}=\left[ -\frac{1}{2}\left( \frac{\partial
^{2}}{\partial x^{2}}+\frac{\partial ^{2}}{\partial y^{2}}\right) -g|\psi
|^{2}+U(x,y)-\Omega \hat{L}_{z}\right] \psi .  \label{GPE}
\end{equation}%
Here, $g\equiv -4\pi \mathcal{N}a_{s}/\left( Ma_{\perp }\right) >0$ is the
effective self-attraction coefficient, with $\mathcal{N}$ the total number
of atoms, $a_{s}<0$ the scattering length of the attractive interatomic
interactions, $a_{\perp }$ the transverse-trapping length, and $M=\int \int
|\psi \left( x,y\right) |^{2}dxdy$ the norm of the 2D wave function, which
is a dynamical invariant of Eq. (\ref{GPE}). The orbital-momentum operator
is $\hat{L}_{z}=i(y\partial _{x}-x\partial _{y})$. In addition to $M$, Eq. (%
\ref{GPE}) conserves the energy,
\begin{equation}
E=\int \left( \frac{1}{2}|\nabla \psi |^{2}-\frac{g}{2}|\psi |^{4}+U|\psi
|^{2}-\Omega \psi ^{\ast }\hat{L}_{z}\psi \right) d\mathbf{r},  \label{E}
\end{equation}%
and the total angular momentum, $L=\int \psi ^{\ast }\hat{L}_{z}\psi d$%
\textbf{$r$}, if the potential is axisymmetric, $U=U(r)$, with $r^{2}\equiv
x^{2}+y^{2}$.

Steady-state solutions to Eq. (\ref{GPE}) are looked for in the ordinary
form, $\psi (x,y,t)=e^{-i\mu t}\phi \left( x,y\right) $, where $\mu $ is the
real chemical potential, while stationary wave function $\phi (x,y)$ remains
complex. To find configurations realizing a minimum of the energy, we used a
modification of the known numerical method based on the integration of the
GPE in imaginary time \cite{Tosi} (in Ref. \cite{rf:1}, a similar method was
used to generate vortex lattices in a 2D model with the repulsive
nonlinearity and harmonic confining potential). To this end, we substitute
real time $t$ in Eq. (\ref{GPE}) by $-i\tau $, introduce additional real
variable $\tilde{M}(\tau )$, and replace Eq. (\ref{GPE}) by the following
system of the Ginzburg-Landau (GL) type:
\begin{eqnarray}
\frac{\partial \phi }{\partial \tau } &=&\left[ \frac{1}{2}\nabla
^{2}+g|\phi |^{2}-U(x,y)+\Omega \hat{L}_{z}+\gamma _{1}\left( \tilde{M}%
-M\right) \right] \phi ,  \label{tau2} \\
\frac{d\tilde{M}}{d\tau } &=&\gamma _{2}\left( M_{0}-M\right) .  \label{M2}
\end{eqnarray}%
Here, $M$ is the same 2D norm as defined above, but it is not a dynamical
invariant of the GL equations, and $M_{0}$ is the target constant value of
the norm in the stationary state to be found, while $\gamma _{1}$ and $%
\gamma _{2}$ are auxiliary positive constants.

Equation (\ref{M2}) acts as a negative feedback, which provides for the
relaxation of the variable norm, $M(\tau )$ , to $M_{0}$. Obviously,
stationary states, into which solutions to Eqs. (\ref{tau2}) and (\ref{M2})
relax at $\tau \rightarrow \infty $, also yield stationary solutions to GPE (%
\ref{GPE}), with chemical potential $\mu =\gamma _{1}\left[ \tilde{M}\left(
\tau \rightarrow \infty \right) -M_{0}\right] $.

Coupled equations (\ref{tau2}) and (\ref{M2}) can be presented in the
gradient form, as
\begin{equation}
\frac{\partial \phi }{\partial \tau }=-\frac{\delta \tilde{E}}{\delta \phi
^{\ast }},\;\;\frac{d\tilde{M}}{d\tau }=\frac{\gamma _{2}}{\gamma _{1}}\frac{%
\partial \tilde{E}}{\partial \tilde{M}},  \label{lag}
\end{equation}%
where $\tilde{E}=E-(1/2)\gamma _{1}(\tilde{M}-M_{0})^{2}+(1/2)\gamma _{1}(M-%
\tilde{M})^{2}$. This representation demonstrates that Eqs. (\ref{tau2}) and
(\ref{M2}) may be regarded as equations generated by the minimization of
functional $E$, under the constraint (Lagrangian condition) that the total
norm is fixed, $M=M_{0}$. Fast convergence of numerical solutions of Eqs. (%
\ref{tau2}) and (\ref{M2}) was achieved, for instance, with the choice of
auxiliary parameters $\gamma _{1}=3$ and $\gamma _{2}=5$. On the other hand,
if the convergence is achieved, the results do not depend on the choice of $%
\gamma _{1}$ and $\gamma _{2}$, which was verified by varying these constant
in a broad range. The numerical integration was performed by means of the
split-step 2D-Fourier method with $256\times 256$ modes.

\section{Lowest-Landau-level states}

\subsection{Quiescent modes}

In the critical case when the confining potential is harmonic, with the
respective frequency equal to the rotation velocity, $U=\left( 1/2\right)
\Omega r^{2}$, Eq.~(\ref{GPE}) and respective energy (\ref{E}) reduce to
\begin{equation}
i\frac{\partial \psi }{\partial t}=\left[ \frac{1}{2}\left( i\frac{\partial
}{\partial x}-\Omega y\right) ^{2}+\frac{1}{2}\left( i\frac{\partial }{%
\partial y}+\Omega x\right) ^{2}-g|\psi |^{2}\right] \psi ,  \label{LLL}
\end{equation}%
\begin{equation}
E=\int \left( \frac{1}{2}|(\nabla -i\mathbf{A})\psi |^{2}-\frac{g}{2}|\psi
|^{4}\right) d\mathbf{r},  \label{E2}
\end{equation}%
where $\mathbf{A}=(-\Omega y,\Omega x)$. As said above, this system with $g=0
$ is equivalent to the 2D Schr\"{o}dinger equation for a charged particle in
uniform magnetic field $\Omega $ directed perpendicular to plane $\left(
x,y\right) $. The wave function of the corresponding ground state, i.e., the
lowest Landau level (LLL), is
\begin{equation}
\psi =A~e^{-i\Omega t}\exp \left[ -\left( \Omega /2\right) \left(
(x-x_{0})^{2}+(y-y_{0})^{2}\right) \right] ,  \label{Landau}
\end{equation}%
where $x_{0}$ and $y_{0}$ determine an arbitrary central position of the
particle, and $A$ is an arbitrary amplitude.

For $g>0$, a localized (quasi-soliton) solution to nonlinear equation (\ref%
{LLL}) may be approximated as an \textit{ansatz} suggested by exact solution
(\ref{Landau}) for the linear equation,
\begin{equation}
\psi =A~e^{-i\mu t}\exp \left[ -\alpha \left(
(x-x_{0})^{2}+(y-y_{0})^{2}\right) \right] ,  \label{ans}
\end{equation}%
where $A$, $\alpha $ and $\mu $ are treated as variational parameters.
Earlier, an ansatz using a product of the LLL wave function and an
appropriate function in the perpendicular direction was used in Ref. \cite%
{Stavros} for the description of 3D vortices in rotating confined BECs.

The norm of ansatz (\ref{ans}) is
\begin{equation}
M=A^{2}\pi /(2\alpha ).  \label{norm}
\end{equation}%
Using this expression, we eliminate $A$ in favor of $M$, and then calculate
energy (\ref{E2}) corresponding to the ansatz:
\begin{equation}
E=M\left[ \alpha +\frac{1}{2}\Omega ^{2}(x_{0}^{2}+y_{0}^{2})+\frac{\Omega
^{2}}{4\alpha }-\frac{gM\alpha }{2\pi }\right] ,  \label{E3}
\end{equation}%
which does not contain $\mu $. To predict the inverse-width parameter $%
\alpha $ of the localized state, we minimize the energy with respect to $%
\alpha $, by setting $\partial E/\partial \alpha =0$. This yields
\begin{equation}
\alpha =\frac{\Omega }{2}\frac{1}{\sqrt{1-gM/(2\pi )}}.  \label{Varalpha}
\end{equation}%
Then, the substitution of this result into Eq. (\ref{norm}) yields the
respective expression for the amplitude,%
\begin{equation}
A^{2}=\frac{\Omega M}{\pi \sqrt{1-gM/(2\pi )}}.  \label{al}
\end{equation}%
Note that expression (\ref{al}) diverges at $gM=2\pi $, which implies the
collapse in the 2D setting due to the self-attraction. This collapse
threshold is a well-known prediction of the VA \cite{Anderson}, which does
not depend on the presence of the external potential or rotation.

Figure 1(a) displays a numerically found profile of the central cross
section of $\psi (x,y)$ for $g=1$ and $M=5$. The dashed curve in the same
figure is the Gaussian fitting to the numerical profile, $\psi _{\mathrm{fit}%
}(x,y)=A_{\mathrm{fit}}\exp \left[ -\alpha _{\mathrm{fit}}(x-L/2)^{2}\right]
$, with $A_{\mathrm{fit}}=1.62$ and $\alpha _{\mathrm{fit}}=0.824$. Note
that the central part of the 2D mode is slightly narrower than the Gaussian,
because $M$ is rather large, inducing self-compression of the mode. The
solid curve in Fig. 1(b) displays the prediction of the VA for inverse-width
parameter $\alpha $, given by Eq. (\ref{Varalpha}), while the chain of
circles represent numerical values of the same parameter, which were found
from an integral expression, $M/\left[ 2\int \int (\left( x-x_{0}\right)
^{2}+\left( y-y_{0}\right) ^{2})|\psi (x,y)|^{2}dxdy\right] $. Indeed, if
ansatz (\ref{ans}) is substituted into this expression, it will yield
exactly $\alpha $.
\begin{figure}[tbp]
\includegraphics[height=4.5cm]{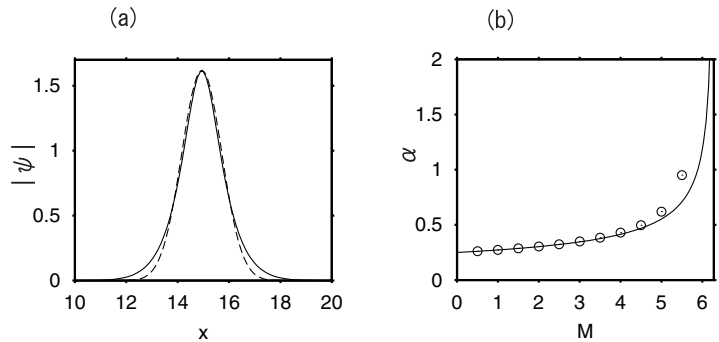}
\caption{(a) A typical shape of the nonlinear localized mode supported by
the LLL (lowest Landau level) for $M=5$. (b) Comparison of the numerically
found inverse-width parameter $\protect\alpha $ with the approximation
provided by the variational method. Other parameters in this figure are $g=1$
and $\Omega =0.5$.}
\label{fig1}
\end{figure}

Modes with the intrinsic angular momentum, i.e., localized vortices, can
also be constructed in this setting. However, a detailed analysis of the
vortex dynamics is beyond the scope of the present work.

\subsection{The Larmor motion}

If the localized state placed at $(x_{0},y_{0})=(0,0)$ is kicked by lending
it wavenumber $k_{y}$ in the $y$-direction, $\psi (x,y)\rightarrow \psi
(x,y)\exp (ik_{y}y)$, the quasi-soliton exhibits rotating Larmor (cyclotron)
motion, as shown in Fig.~2(a). If a ramp (constant external force) is
applied in the $x$-direction, by adding the extra potential term, $U_{%
\mathrm{extra}}(x)=-Fx$, to energy (\ref{E2}), the quasi-soliton gets
engaged in a drift motion, as shown in Fig.~2(b). These scenarios of the
Larmor motion of the quasi-soliton are qualitatively the same as exhibited
by the LLL localized state in the limit of the linear Schr\"{o}dinger
equation ($g=0$), as shown in Fig.~2(c). One can check, by means of direct
simulations, that a localized vortex with vorticity $m=1$ also exhibits the
Larmor motion in the same linear Schr\"{o}dinger equation.
\begin{figure}[tbp]
\includegraphics[height=3.5cm]{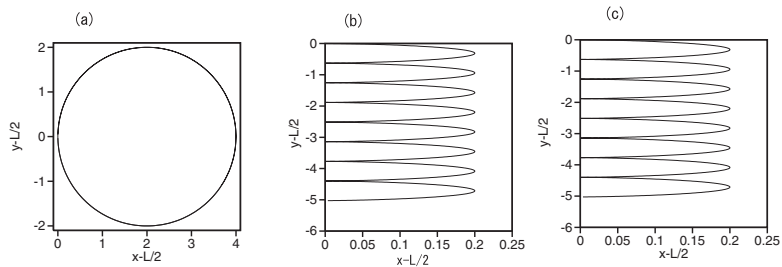}
\caption{(a) The Larmor motion of the c.m. of the LLL quasi-soliton
initiated by the application of kick $k_{y}=2$ in the $y$-direction, for $g=1
$ and $M=5$. (b) The drift of the c.m. in the ramp potential, $U_{\mathrm{%
extra}}=-0.1x$, again for $g=1$ and $M=5$. (c) The drift of the c.m. under
the action of the same tilted potential, but for for $g=0$ (in the linear
Schr\"{o}dinger equation).}
\label{fig2}
\end{figure}

The Larmor motion induced by the additional potential can also be explained
via the variational method. The Lagrangian corresponding to Eq. (\ref{LLL}),
with the addition of the extra potential, is
\begin{equation}
L=\int \left( \frac{i}{2}(\psi _{t}\psi ^{\ast }-\psi _{t}^{\ast }\psi )-%
\frac{1}{2}|(\nabla -i\mathbf{A})\psi |^{2}+\frac{g}{2}|\psi |^{4}-U_{%
\mathrm{extra}}(x,y)|\psi |^{2}\right) d\mathbf{r}.  \label{La}
\end{equation}%
If $\psi (x,y,t)$ is approximated by a generalization of ansatz (\ref{ans}),
\textit{viz}.,
\begin{eqnarray}
\psi &=&A~e^{-i\mu t}\exp \left[ -\alpha \left(
(x-x_{0}(t))^{2}+(y-y_{0}(t))^{2}\right) \right]  \notag \\
&&\times \exp [ip_{x}(x-x_{0}(t))+ip_{y}(y-y_{0}(t))],  \label{ans2}
\end{eqnarray}%
the corresponding effective Lagrangian takes the following form, after a
straightforward calculation:
\begin{eqnarray}
\frac{L_{\mathrm{eff}}}{M} &=&p_{x}\dot{x}_{0}+p_{y}\dot{y}_{0}-\alpha -%
\frac{\Omega ^{2}}{2}(x_{0}^{2}+y_{0}^{2})-\frac{1}{2}(p_{x}^{2}+p_{y}^{2})
\notag \\
&&-\frac{1}{4\alpha }\Omega ^{2}+\frac{gM\alpha }{2\pi }-\Omega
(p_{x}y_{0}-p_{y}x_{0})-U_{\mathrm{eff}}\left( x_{0},y_{0}\right) ,
\label{Leff}
\end{eqnarray}%
where the overdot stands for the time derivative, and
\begin{equation}
U_{\mathrm{eff}}(x_{0},y_{0},\alpha )\equiv \frac{A^{2}}{M}\int \int U_{%
\mathrm{extra}}(x,y)\exp \left[ -\alpha \left(
(x-x_{0}(t))^{2}+(y-y_{0}(t))^{2}\right) \right] dxdy.  \label{Ueff}
\end{equation}%
The system of the Euler-Lagrange variational equations is then derived from
the effective Lagrangian:%
\begin{eqnarray}
\frac{dp_{x}}{dt} &=&\Omega p_{y}-\frac{\partial U_{\mathrm{eff}}}{\partial
x_{0}},  \notag \\
\frac{dp_{y}}{dt} &=&-\Omega p_{x}-\frac{\partial U_{\mathrm{eff}}}{\partial
y_{0}},  \notag \\
\frac{dx_{0}}{dt} &=&p_{x}+\Omega y_{0},  \label{motion1} \\
\frac{dy_{0}}{dt} &=&p_{y}-\Omega x_{0},  \notag \\
-1 &+&\frac{gM}{2\pi }+\frac{\Omega ^{2}}{4\alpha ^{2}}-\frac{\partial U_{%
\mathrm{eff}}}{\partial \alpha }=0.  \notag
\end{eqnarray}

In the simplest case of the uniform ramp, corresponding to $U_{\mathrm{extra}%
}=-Fx$, i.e., $\partial U_{\mathrm{eff}}/\partial \alpha =0$, the last
equation in system (\ref{motion1}) yields the same constant expression for $%
\alpha $ as given above by Eq. (\ref{Varalpha}). With constant $\alpha $
(i.e., constant width of the LLL mode), the remaining part of system (\ref%
{motion1}) amounts to coupled equations of motion of the second order,
\begin{eqnarray}
\frac{d^{2}x_{0}}{dt^{2}} &=&2\Omega \frac{dy_{0}}{dt}-\frac{\partial U_{%
\mathrm{eff}}}{\partial x_{0}},  \notag \\
&&  \label{motion2} \\
\frac{d^{2}y_{0}}{dt^{2}} &=&-2\Omega \frac{dx_{0}}{dt}-\frac{\partial U_{%
\mathrm{eff}}}{\partial y_{0}}.  \notag
\end{eqnarray}%
With $U_{\mathrm{eff}}=0$, this system describes the Larmor (cyclotron)
motion, and for $U_{\mathrm{eff}}=-Fx_{0}$, Eqs. (\ref{motion2}) predict an
overlap of the Larmor rotation and drift in the $y$-direction. The drift
motion, in the direction perpendicular to the ramp, is an analog of the
ordinary Hall effect in solid-state physics (similarities between the
soliton dynamics in BEC and the quantum Hall effect were discussed in
various contexts, see, e.g., Ref. \cite{Hall} and references therein).
Additional analogies to the Hall effect are considered below.

If two LLL quasi-solitons are kicked in opposite directions, by
imparting wavenumbers $\pm k_{y}$ to them, the Larmor motion of
the solitons eventually results in a collision. If the norm of
each quasi-soliton is not too large, the collision seems
completely elastic, and the two objects keep moving in closed
trajectories which together form a ``figure
of eight", surviving multiple collisions, as shown in Fig. 3(a) for $M=1$, $%
g=1$ and $k_{y}=2$. On the other hand, if the norm of each LLL
mode is larger, the collision is inelastic, reducing the
wavenumbers of the quasi-solitons, and thus making the radius of
the Larmor motion smaller. This case is illustrated by Fig. 3(b)
for $M=5$. Because the respective total norm,
$M_{\mathrm{tot}}=10$, exceeds the collapse threshold, the two
quasi-solitons merge into a single collapsing object, after
several consecutive collisions. Inelastic collisions between 2D
solitons which are too ``heavy" and can also suffer the merger and
collapse
were reported in different settings, such as quasi-1D guiding channels \cite%
{BBB}.
\begin{figure}[tbp]
\includegraphics[height=5.5cm]{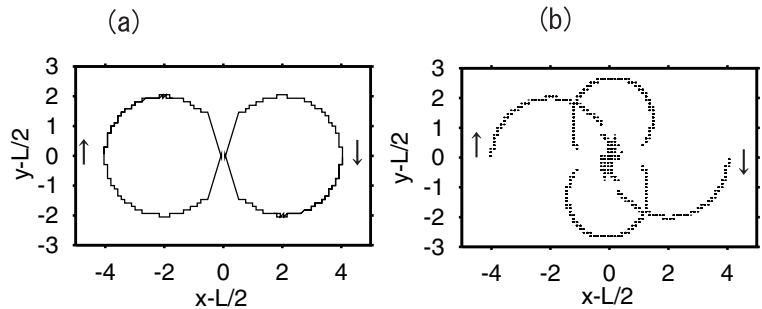}
\caption{(a) Trajectories of points of maxima of $\left\vert \protect\psi %
(x,y)\right\vert $ in half-planes $x>L/2$ and $x<L/2$ for a pair of LLL
quasi-solitons with norm $M=1$, kicked by $k_{y}=\pm 2$. The solitons
survive multiple elastic solutions. The arrows shows the initial positions
and velocities of the two solitons. (b) The same for $M=5$. In this case,
the collision results in the fusion and collapse of the LLL quasi-solitons.}
\label{fig3}
\end{figure}

\subsection{The Hall effect and edge states for the quasi-solitons}

As mentioned above, the analog of the Hall effect for matter-wave solitons
is a subject of considerable interest \cite{Hall}. To study it in the
present context, i.e., as a matter of fact, to consider the corresponding
\textit{edge states} of the LLL quasi-solitons, we combine the ramp with a
weak holding quartic potential in the $x$-direction, by taking $U_{\mathrm{%
extra}}=-Fx+bx^{4}$, with small $b>0$.

The VA can be used in this setting too. To this end, we approximate $\psi $
by an anisotropic Gaussian ansatz,
\begin{eqnarray}
\psi &=&A~e^{-i\mu t}\exp [-\alpha (x-x_{0})^{2}-\beta (y-y_{0})^{2}]  \notag
\\
&&\exp [ip_{x}(x-x_{0}(t))+ip_{y}(y-y_{0}(t))],  \label{aniso}
\end{eqnarray}%
cf. expression (\ref{ans2}). The substitution of ansatz (\ref{aniso}) in
Lagrangian (\ref{La}) yields
\begin{eqnarray}
\frac{L_{\mathrm{eff}}}{M} &=&p_{x}\dot{x}_{0}+p_{y}\dot{y}_{0}-\frac{1}{2}%
(\alpha +\beta )-\frac{\Omega ^{2}}{2}(x_{0}^{2}+y_{0}^{2})-\frac{1}{2}%
(p_{x}^{2}+p_{y}^{2})  \notag \\
&&-\frac{\Omega ^{2}}{8}\left( \frac{1}{\alpha }+\frac{1}{\beta }\right) +%
\frac{gM}{2\pi }\sqrt{\alpha \beta }-\Omega (p_{x}y_{0}-p_{y}x_{0})-U_{%
\mathrm{eff}}\left( x_{0},y_{0}\right) ,  \label{Leff2}
\end{eqnarray}%
where, this time, the norm is $M=A^{2}\pi /\left( 2\sqrt{\alpha \beta }%
\right) $, and
\begin{equation}
U_{\mathrm{eff}}\left( x_{0},y_{0}\right)
=-Fx_{0}+b(x_{0}^{2}+y_{0}^{2})^{2}+\frac{b}{2\alpha }(3x_{0}^{2}+y_{0}^{2})+%
\frac{b}{2\beta }(x_{0}^{2}+3y_{0}^{2}).
\end{equation}%
cf. Eqs. (\ref{Leff}) and (\ref{Ueff}). The respective equations of motion
are written as
\begin{eqnarray}
\frac{d^{2}x_{0}}{dt^{2}} &=&2\Omega \frac{dy_{0}}{dt}-\frac{\partial U_{%
\mathrm{eff}}}{\partial x_{0}},  \notag \\
\frac{d^{2}y_{0}}{dt^{2}} &=&-2\Omega \frac{dx_{0}}{dt}-\frac{\partial U_{%
\mathrm{eff}}}{\partial y_{0}},  \notag \\
&&  \label{Lag2} \\
-\frac{1}{2} &+&\frac{gM}{4\pi }\sqrt{\frac{\beta }{\alpha }}+\frac{\Omega
^{2}}{8\alpha ^{2}}-\frac{\partial U_{\mathrm{eff}}}{\partial \alpha }=0,
\notag \\
-\frac{1}{2} &+&\frac{gM}{4\pi }\sqrt{\frac{\alpha }{\beta }}+\frac{\Omega
^{2}}{8\beta ^{2}}-\frac{\partial U_{\mathrm{eff}}}{\partial \beta }=0.
\notag
\end{eqnarray}%
Stationary solutions to Eqs. (\ref{Lag2}) are determined by algebraic
relations,
\begin{equation}
\dot{y}_{0}=0,\;4bx_{0}^{3}+b(3/\alpha +1/\beta )x_{0}=F
\end{equation}%
\begin{equation}
-(1/2)+gM/(4\pi )\sqrt{\beta /\alpha }+\Omega ^{2}/(8\alpha
^{2})+3bx_{0}^{2}/(2\alpha ^{2})=0,  \label{algebra}
\end{equation}%
\begin{equation}
-(1/2)+gM/(4\pi )\sqrt{\alpha /\beta }+\Omega ^{2}/(8\beta
^{2})+bx_{0}^{2}/(2\beta ^{2})=0.
\end{equation}%
In the linear limit ($g=0$), these relations yield $\alpha =\sqrt{\Omega
^{2}/4+3bx_{0}^{2}}$ and $\beta =\sqrt{\Omega ^{2}/4+bx_{0}^{2}}$, with
wavenumber in the $y$-direction being $p_{y}=\Omega x_{0}>0$.

As an example, we refer to the LLL mode in the edge state, that was found in
the numerical form for $M=5$, $F=0.5$, $b=0.0005$, $g=1$, and $\Omega =0.5$%
.\ Fitting this mode to Gaussian ansatz (\ref{aniso}), the corresponding
parameters were numerically evaluated as $x_{0}=6.225,\alpha =0.724,\beta
=0.679$, $A\equiv (4\alpha \beta /\pi )^{1/4}\sqrt{M}=1.99$, and $%
p_{y}=\Omega x_{0}$. Figures 4(a) and 4(b) display, respectively, the
contour plot of \textrm{Re}$\left\{ \psi \left( x,y\right) \right\} $ and
cross-section profile $|\psi (x,y)|$ at $y=L/2$, which is compared to its
counterpart predicted by the VA through Eqs. (\ref{aniso}) and (\ref{algebra}%
). In addition, Fig. 4(c) displays $\mathrm{Re}\left\{ \psi (x,y)\right\} $
at $x=21.2$, which is compared to the respective approximation provided by
ansatz (\ref{aniso}), $A\exp \{-\beta (y-L/2)^{2}\}\cos (p_{y}y)$.
\begin{figure}[tbp]
\includegraphics[height=4.5cm]{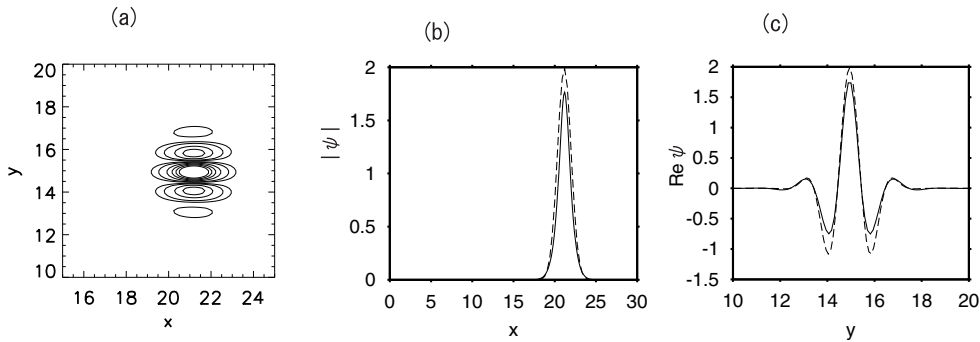}
\caption{(a) The contour plot of $\mathrm{Re}\left\{ \protect\psi %
(x,y)\right\} $ of the edge state, for $g=1$, $M=5$,$~F=0.5$ and $b=0.0005$.
(b) The profile of $|\protect\psi (x,y)|$ at $y=L/2$. (c) $\mathrm{Re}%
\left\{ \protect\psi (x,y)\right\} $ at $x=21.2$. The two latter panels
include comparison with profiles predicted by the variational approximation.}
\label{fig4}
\end{figure}

If the edge state is kicked with wavenumber $k_{y}$, the quasi-soliton
exhibits drift motion. For parameters identical to those in Fig. 4, in Fig.
5 we display the trajectory of its c.m., initiated by the kick with $k_{y}=1$%
, and the same trajectory as predicted by variational equations~(\ref{Lag2}%
). These results demonstrate that the VA provides quite a reasonable
accuracy for the description of dynamical states, as well as for static
ones.
\begin{figure}[tbp]
\includegraphics[height=4.5cm]{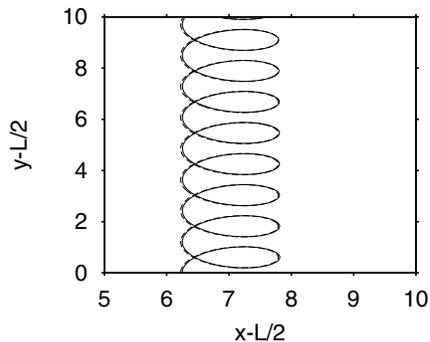}
\caption{The trajectory of the drift motion of an edge LLL mode, kicked with
$k_{y}=1$, for $g=1$, $F=0.5$, $b=0.0005$, and $M=5$, as obtained from
direct numerical simulations of the GPE (the solid curve), and the
counterpart of the same trajectory, predicted by the variational
approximation (the dashed curve).}
\label{fig5}
\end{figure}

\section{The axisymmetric quartic potential}

\subsection{Quiescent states: vortices, quasi-solitons, and crescents}

In the rest of the paper, we focus on the situation opposite to that
considered in the previous section, namely, the setting with the quartic
axisymmetric potential, $U(x,y)=br^{4}$, where $b$ is a small positive
constant. The purpose of the consideration of this model is to highlight the
dynamics on vortices and quasi-solitons under the anharmonic confinement.
The respective variant of the GPE is [cf. Eq.~(\ref{LLL})]%
\begin{equation}
i\frac{\partial \psi }{\partial \tau }=\left[ -\frac{1}{2}\nabla ^{2}-\Omega
\hat{L}_{z}-g|\phi |^{2}+\frac{b}{2}r^{4}\right] \phi ,  \label{quartic}
\end{equation}%
where $\nabla ^{2}$ is the 2D Laplacian, and the angular-momentum operator
is $\hat{L}_{z}=i\left( x\partial _{y}-y\partial _{x}\right) $.

Similar to Eqs. (\ref{tau2}) and (\ref{M2}), steady-state solutions to Eq. (%
\ref{GPE}) are looked for through the simulations of relaxation in the
following modified system of the GL type:
\begin{eqnarray}
\frac{\partial \phi }{\partial \tau } &=&\left[ \frac{1}{2}\nabla
^{2}+g|\phi |^{2}-\frac{b}{2}r^{4}+\Omega \hat{L}_{z}+\gamma _{1}\left(
\tilde{M}-M\right) \right] \phi ,  \label{tau} \\
\frac{d\tilde{M}}{d\tau } &=&\gamma _{2}\left( M_{0}-M\right) .  \label{M}
\end{eqnarray}%
Here, $M$ is the same 2D norm as defined above, and $M_{0}$ is the target
constant value of the norm in the stationary state to be found.

Using the remaining scaling invariance of Eq. (\ref{GPE}), we fix
normalizations by choosing $b=0.002$ (a small value of trapping coefficient $%
b$ is necessary to allow the condensate enough room to evolve), and $M_{0}=4$%
. Axisymmetric vortex solutions to Eqs. (\ref{tau}) and (\ref{M}) with
topological charge $m$ are sought for as $\phi =r^{m}e^{im\theta
}R_{m}(r,\tau )$, where the reduced amplitude function, $R_{m}$, satisfies
the following equations:
\begin{equation}
\frac{\partial R_{m}}{\partial \tau }=\frac{1}{2}\left( \frac{\partial ^{2}}{%
\partial r^{2}}+\frac{2m+1}{r}\frac{\partial }{\partial r}-br^{4}\right)
R_{m}+gr^{2m}R_{m}^{3}+m\Omega R_{m}+\alpha (\tilde{M}-M)R_{m},  \label{R}
\end{equation}
\begin{equation}
\frac{d\tilde{M}}{d\tau }=\beta (M_{0}-M),~M(\tau )=2\pi \int_{0}^{\infty
}r^{2m+1}R_{m}^{2}dr.  \label{RM1}
\end{equation}%
Once stationary solutions were found, their stability was examined by means
of computation of the respective eigenvalues, using the Bogoliubov - de
Gennes equations, i.e., the linearization of Eq. (\ref{GPE}), for small
perturbations built (for given $m$) as superpositions of components with
vorticities $m\pm 1$ (such perturbation modes turn out to be most dangerous
in terms of the instability).

The results, in the form of a stability diagram in the plane of $\left(
\Omega ,g\right) $ for $m=0,1,2,3,4,5$, are presented in Fig.~6(a). Each
stability domain for $m\geq 1$ is bounded by two curves, which are generated
by critical perturbation eigenmodes that are found to be, respectively, real
and imaginary, with respect to unperturbed amplitude functions $R_{m}(r)$.
This stability diagram is qualitatively similar to those reported in the
model with a mixed quadratic-quartic radial trap \cite{mean-field,many-body}%
. 
\begin{figure}[tbp]
\includegraphics[height=4.5cm]{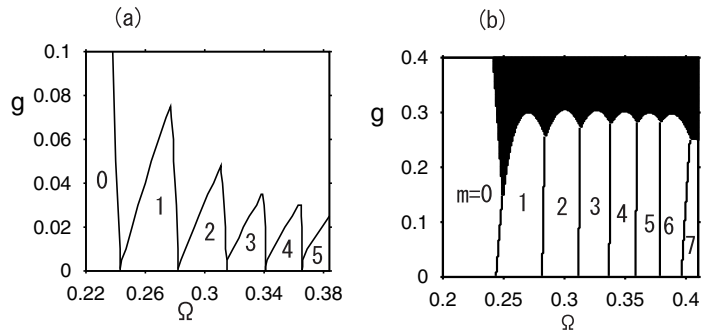}
\caption{(a) The numerically generated stability diagram for vortices with
different values of topological charge $m$, in the plane of the rotation
frequency and strength of the attractive interaction, $\Omega $ and $g$, in
the model with the quartic confining potential. (b) Regions where the
variational approximation predicts that vortices with charge $m$ or c.m.
states, alias quasi-solitons (the black area), provide for a minimum of the
energy. }
\label{fig6}
\end{figure}
For given $\Omega >0$, vortices with $m<0$ can be found too, but they all
are unstable. This instability can be readily explained by the fact that the
Coriolis term in the vortex' energy, which is proportional to $-m\Omega $,
is positive for $m<0$, see Eq. (\ref{Evort}) below.

\begin{figure}[tbp]
\includegraphics[height=4.cm]{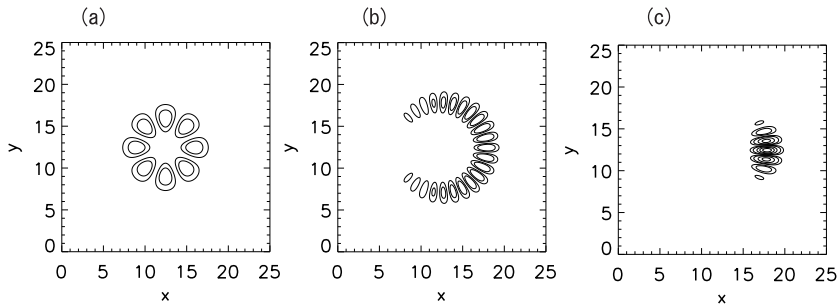}
\caption{Generic examples of stable matter-wave patterns in the rotating
condensate confined by the quartic potential are shown by means of contour
plots of $|\mathrm{Re}\{\protect\phi (x,y)\}|$: (a) a vortex with $m=4$, for
$g=0.01,~\Omega =0.35$; (b) a crescent, for $g=0.001,~\Omega =0.5$; (c) a
strongly compressed c.m. state (quasi-soliton), for $g=0.5,~\Omega =0.5$.}
\label{fig7}
\end{figure}

While the increase of $\Omega $ at fixed self-attraction coefficient $g$
leads to the transition to vortices with larger $m$, the vortex states
become unstable with the increase of $g$, being replaced by crescent-shaped
ones, as shown in Fig. \ref{fig7}. In Refs. \cite{mean-field} and \cite{He},
similar patterns were interpreted (in models with the local and nonlocal
nonlinearity, respectively) as superpositions of vortices with three
different values of the topological charge, \textit{viz}., given $m$ and $%
m\pm 1$. This fact agrees with the above-mentioned finding that perturbation
modes which can destabilize a given vortex carry vorticities $m\pm 1$.
Further increase of $g$ leads to a reduction of the crescent's length and
its gradual compression into a strongly localized c.m. state, alias
quasi-soliton, which carries an intrinsic phase gradient along the azimuthal
direction, see Fig. \ref{fig7}(c); note that the value of $g$ corresponding
to the quasi-soliton falls below the collapse threshold. The shift of the
c.m. state from the rotation pivot increases with $\Omega $. This
quasi-soliton is similar to the edge state of the LLL type shown above in
Fig.~4.

In the limit of large $m$, the stationary version of Eqs. (\ref{R}) and (\ref%
{RM1}) give rise to a simple \emph{asymptotically exact} solution, which
does not depend on $g$ and $b$:
\begin{equation}
R_{m}(r)=\sqrt{M_{0}\Omega ^{m+1}/\left( \pi m!\right) }\exp \left( -\Omega
r^{2}/2\right)  \label{asympt}
\end{equation}%
(recall we here fix the norm as $M_{0}=4$). This solution, and numerical
results obtained for finite $m$, suggest to approximate the stationary
solution for vortices by ansatz $\phi =Ae^{im\theta }r^{m}\exp \left(
-\alpha r^{2}\right) $, whose norm is $M_{0}=\pi m!\left( 2\alpha \right)
^{-(m+1)}A^{2}$. The substitution of the ansatz in Eq. (\ref{E}) yields the
corresponding expression for the energy,
\begin{equation}
\frac{E}{M_{0}}=-m\Omega +\frac{(m+2)(m+1)b}{4\alpha ^{2}}+\alpha
(m+1)-(2m)!M_{0}g\alpha /\left[ 2^{2m+1}\left( m!\right) ^{2}\pi \right] .
\label{Evort}
\end{equation}%
Then, width parameter $\alpha $ for the solution sought for is determined by
the minimization of the energy, $\partial E/\partial \alpha =0$, which
yields
\begin{equation}
\alpha ^{-3}=\frac{2}{b(m+2)(m+1)}\left[ m+1-\frac{(2m)!M_{0}g}{%
2^{2m+1}(m!)^{2}\pi }\right] .  \label{alpha}
\end{equation}

On the other hand, a quasi-soliton with the c.m. located at distance $x_{0}$
from the pivot, see Fig. \ref{fig7}(c), may be approximated by the
anisotropic ansatz, which is similar to the one used above in Eq. (\ref%
{aniso}),
\begin{equation}
\psi =A\exp \left[ iky-\left( \alpha (x-x_{0})^{2}+\beta y^{2}\right) \right]
,  \label{quasi}
\end{equation}%
with norm $M_{0}=\pi A^{2}/\left( 2\sqrt{\alpha \beta }\right) $. If this
c.m. state was generated by an unstable vortex with charge $m$, comparison
of the azimuthal phase gradients suggests that $k=m/x_{0}$. When substituted
in Eq. (\ref{E}), ansatz (\ref{quasi}) yields
\begin{equation}
\frac{E}{M_{0}}=\frac{\alpha +\beta +k^{2}}{2}-\Omega kx_{0}-\frac{g\sqrt{%
\alpha \beta }M_{0}}{2\pi }+\frac{b\left[ 3\beta ^{2}+2\alpha \beta
(1+12\beta x_{0}^{2})+\alpha ^{2}(3+8\beta x_{0}^{2}+16\beta ^{2}x_{0}^{4})%
\right] }{16\alpha ^{2}\beta ^{2}}.  \label{Esol}
\end{equation}%
Values of the variational parameters are predicted by equations $\partial
E/\partial x_{0}=\partial E/\partial k=\partial E/\partial \alpha =\partial
E/\partial \beta =0$, which yield, in particular,%
\begin{equation}
k=\Omega x_{0},~x_{0}^{2}=\left[ \Omega ^{2}\alpha \beta -b(3\beta +\alpha )%
\right] /\left( 4b\alpha \beta \right) ,  \label{x0}
\end{equation}%
if $x_{0}\neq 0$. Another solution, with $x_{0}=0$ (an isotropic soliton
sitting at the center, which is stable for small $\Omega $) has $k=0$ and $%
\alpha =\beta =\left[ (2\pi b)/(2\pi -M_{0})\right] ^{1/3}$. Further
analysis of the VA solutions (without fixing $M_{0}=4$) demonstrates that
they predict the collapse (nonexistence of solutions) at $M_{0}\geq 2\pi $,
in accordance with the known variational result \cite{Anderson}.

Using the VA solutions and expressions (\ref{Evort}) and (\ref{Esol}), we
have identified, as shown in Fig. \ref{fig6}(b), regions in parameter plane $%
\left( \Omega ,g\right) $ where vortices with particular integer values of $m
$, or the c.m. state provide for the minimum of the energy, i.e., determine
the ground state. Comparison with the numerically identified stability
regions for vortices in Fig. \ref{fig6}(a) demonstrates that the VA predicts
transitions between different values of $m$ with the increase of $\Omega $
quite accurately. The discrepancy in Fig. \ref{fig6} between the numerical
and variational plots in the direction of $g$ has an obvious reason: the VA
does not take into regard the other species of the localized states, viz.,
crescents, alias mixed-vorticity states [see Fig. \ref{fig7}(b)]. In fact,
crescents have their own domain of the energy dominance, between those of
the vortices and c.m. states. Note that solution (\ref{asympt}), which is
asymptotically exact for $m\rightarrow \infty $, and expression (\ref{Evort}%
) predict the equality between energies of vortices with $m$ and $m+1$,
i.e., borders between the energy-dominance areas of these states (for $%
g\rightarrow 0$), at $\Omega _{m}\approx \left( 4bm\right) ^{1/3}$. To
directly test the accuracy of the VA, in Fig.~8 we display a comparison
between characteristics of the quasi-solitons, \textit{viz}., c.m. offset $%
x_{0}$, amplitude $A$, and wavenumber $k$, as found from numerical results
and predicted by the VA.
\begin{figure}[tbp]
\includegraphics[height=4.cm]{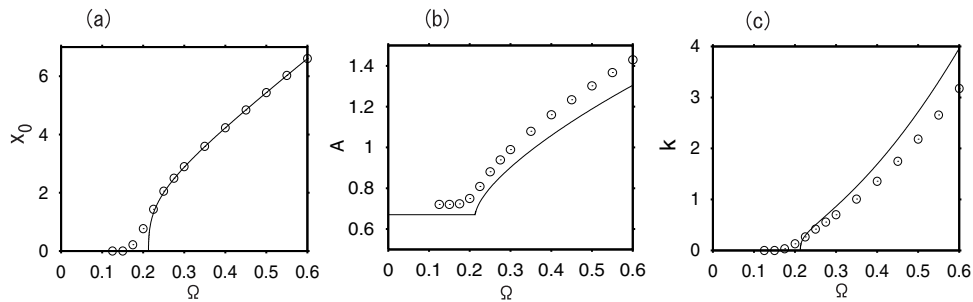}
\caption{Parameters of stable quasi-solitons trapped in the quartic
axisymmetric potential, as found from the numerical computations (circles)
and predicted by the variational approximation (curves) for $g=1$: (a) the
shift from the rotation pivot; (b) amplitude; (c) intrinsic wavenumber.}
\label{fig8}
\end{figure}

\begin{figure}[tbp]
\includegraphics[height=4.cm]{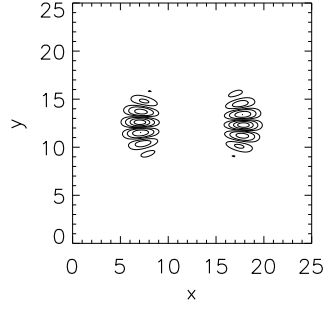}
\caption{A generic example of a stable pair of two strongly compressed c.m.
states found for $g=0.5$, $\Omega =0.5$.}
\label{additional-figure}
\end{figure}

The variational results presented above suggest that the model may also
support a pair of c.m. states placed at diametrically opposite points.
Indeed, Eq. (\ref{x0}) gives rise to two roots, $x_{0}=\pm \sqrt{\left[
\Omega ^{2}\alpha \beta -b(3\beta +\alpha )\right] /\left( 4b\alpha \beta
\right) }$, which correspond to opposite values of $y$-wavenumber $k$. Such
\emph{stable} quasi-soliton pairs can be readily found from the numerical
solution, see a typical example in Fig. \ref{additional-figure}.
In the experiment, the pair can be created, for instance, by originally
adding a strong blue-detuned (repulsive) light sheet which cuts the circular
trap into semi-circles. After two c.m. states have self-trapped, the sheet
may be turned off, to restore the axial symmetry of the trap.

\section{The motion and collisions of quasi-solitons}

\begin{figure}[tbp]
\includegraphics[height=3.5cm]{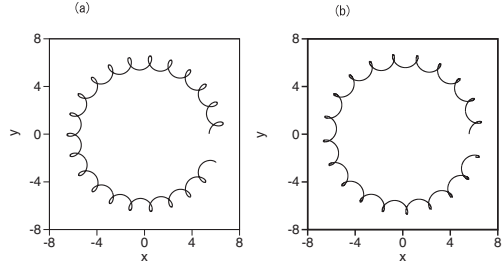}
\caption{(a) The trajectory of the c.m. motion of a quasi-soliton kicked
with wavenumber $q=1$, for $g=1$, $\Omega =0.5$, in the model with the
quartic axisymmetric trap. (b) Its counterpart predicted by the variational
approximation.}
\label{fig9}
\end{figure}

Although they were obtained above as quiescent solutions, the localized c.m.
states can be readily set in motion by the application of tangential kick $%
\exp (iqy)$ with wavenumber $q$, similar to how this was done above for the
LLL localized modes. As a result, the quasi-soliton exhibits rotary motion,
following a trajectory in the form of an \textit{epitrochoid}, see a typical
example in Fig. \ref{fig9}.

Following the approach elaborated above for the moving LLL modes, we present
a similar analytical description of the kicked quasi-soliton in the model
with the quartic confinement. To this end, we adopt the same ansatz (\ref%
{ans2}) as used above, and use the Lagrangian corresponding to Eq. (\ref%
{quartic}),%
\begin{equation}
L=\int \left( \frac{i}{2}(\psi _{t}\psi ^{\ast }-\psi _{t}^{\ast }\psi )-%
\frac{1}{2}|\nabla \psi |^{2}+\Omega \psi ^{\ast }\hat{L}_{z}\psi +\frac{g}{2%
}|\psi |^{4}-\frac{1}{2}br^{4}|\psi |^{2}\right) d\mathbf{r},
\end{equation}%
cf. Eq. (\ref{La}). The substitution of the ansatz in this Lagrangian and
straightforward calculations lead to the following equations of motion for
the c.m. mode:
\begin{eqnarray}
\frac{d^{2}x_{0}}{dt^{2}} &=&2\Omega \frac{dy_{0}}{dt}-\frac{\partial U_{%
\mathrm{eff}}}{\partial x_{0}},  \notag \\
&&  \label{motion3} \\
\frac{d^{2}y_{0}}{dt^{2}} &=&-2\Omega \frac{dx_{0}}{dt}-\frac{\partial U_{%
\mathrm{eff}}}{\partial y_{0}},  \notag
\end{eqnarray}%
with $U_{\mathrm{eff}}(r)=-(1/2)\Omega ^{2}r^{2}+br^{4}+(2b/\alpha )r^{2}$.
Figure \ref{fig9}(b) displays a counterpart of the numerically found
trajectory from panel \ref{fig9}(a), as produced by Eqs. (\ref{motion3}) for
$g=1,b=0.002$, and $\alpha =0.532$. The initial velocity is $%
dx_{0}/dt=0,dy_{0}/dt=1$. It is seen that the VA matches the numerical
findings very well in this case too.

\begin{figure}[tbp]
\includegraphics[height=4cm]{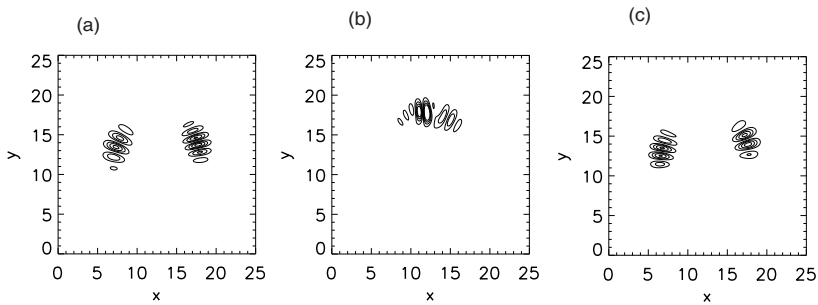}
\caption{Snapshots of $|\mathrm{Re}\left\{ \protect\psi (x,y,t\right\} |$,
at $t=10$ (a), $t=50$ (b), and $t=90$ (c) illustrating an elastic collision
between counter-rotating c.m. states kicked by $q=\pm 0.5$, at $g=0.5$, $%
\Omega =0.5$. }
\label{fig11}
\end{figure}

The possibility of the motion of the c.m. states suggests to consider
collisions between them, also in analogy with what was done above for the
LLL modes. In Fig. \ref{fig11}, we display a typical example of the
collision, which is generated by applying kicks with $q=\pm 0.5$ to
identical quasi-solitons with their c.m. placed, originally, at
diametrically opposite points (in the experiment, such an initial
configuration can be created as outlined at the end of the previous
subsection). The kicked solitons are no longer identical because term $%
\Omega \hat{L}_{z}$ in Eq. (\ref{quartic}) breaks the symmetry between the
clockwise and counter-clockwise directions of the rotation. Nevertheless,
the collisions are elastic, with the quasi-solitons recovering their shapes
after the collision.

\section{Conclusion}

We have revisited the 2D model of rotating BEC with attraction between
atoms. Two different situations of special physical interest were
considered: the one with the critical strength of the quadratic confining
potential, and the purely quartic axisymmetric trap. In the former case, the
linear limit of the GPE (Gross-Pitaevskii equation) is tantamount to the Schr%
\"{o}dinger equation for a charged particle moving in the uniform magnetic
field. We have demonstrated that the action of the self-focusing
nonlinearity on the localized state corresponding to the wave function at
the LLL (lowest Landau level) gives rise to stable quasi-solitons. These
states, both quiescent ones and those set in motion by the kick, or under
the action of the ramp potential, are very accurately described by the VA
(variational approximation). We have also considered the situation when an
external weak 1D quartic potential acts in combination with the ramp, which
gives rise to edge states emulating the Hall effect in terms of the
matter-wave quasi-solitons.

In the case when the axisymmetric trap is represented by the quartic
potential, we have developed the VA which provides for an accurate
description of two species of stable localized states in the model, namely,
vortices with an arbitrary value of the topological charge, and c.m. modes
shifted off the rotation pivot, alias quasi-solitons. Stable states in the
form of two c.m. modes placed at diametrically opposite sites were found
too. The other species, crescents, was obtained in the numerical form. It
was also demonstrated that kicking the c.m. state in the tangential
direction sets it in motion along an \textit{epitrochoidal }trajectory, and
collisions between such solitons are elastic. The motion of the kicked
quasi-soliton in the latter situation is also accurately predicted by the VA.

\end{document}